\documentstyle[aps,preprint]{revtex}

\def\text#1{\mbox{#1}}

\tightenlines
\begin{document}

\title{{A superspace gauge-invariant formulation of a massive tridimensional 2-form field}}
\author{M. A. M. Gomes, R. R. Landim, C. A. S. Almeida\footnote{%
Electronic address:carlos@fisica.ufc.br}}
\address{ Universidade Federal do Cear\'{a}\\
 Physics Department\\
 C.P. 6030, 60470-455 Fortaleza-Ce, Brazil}
\maketitle
\begin{abstract}
By dimensional reduction of a massive supersymmetric B$\wedge $F
theory, a manifestly $N=1$ supersymmetric completion of a massive
antisymmetric tensor gauge theory is constructed in (2+1)
dimensions. In the $N=1-D=3$ superspace, a new topological term is
used to give mass for the Kalb-Ramond field. We have introduced a
massive gauge invariant model using the St\"{u}ckelberg formalism and an
abelian topologically massive theory for the Kalb-Ramond superfield.
An equivalence of both massive models is suggested. Further, a
component field analysis is performed, showing a second
supersymmetry in the model.

\end{abstract}

\section{Introduction}

Antisymmetric tensor fields appear in many field theories. In particular,
the Kalb-Ramond gauge field plays an important role in strong-weak coupling
dualities among string theories \cite{sen} and in axionic cosmic strings
\cite{copeland}. On the other hand, a first order formulation of the
non-Abelian Yang-Mills gauge theory ( BF-YM model) \cite{martellini,fucito}
makes use of a two form gauge potential $B$ to contribute to a discussion of
the problem of quark confinement in continuum QCD \cite{hooft}. Another
interesting aspect of the (3+1) dimensional $B\wedge F$ term ($F=dA$ is the
field strength of a one form gauge potential $A$) is its ability to give
rise to gauge invariant mass to the gauge field \cite{allen}. This property
has been used to obtain an axion field topologically massive and an axionic
charge on a black hole as well \cite{lahiri}. In addition, the existence of
the Higgs mechanism to the Kalb-Ramond gauge fields was demonstrated by
S.-J. Key \cite{key} in the context of closed strings. On the other hand, if
coupled to open strings, the KB\ field becomes a massive vector field
through the St\"{u}ckelberg mechanism. Also, we can mention a topologically
massive Kalb-Ramond field in a $D=3$ context that was introduced in ref.
\cite{dedit}.

It is known that massless string excitations may be described by a
low-energy supergravity theory and that a massless gravity supermultiplet of
graviton, dilaton and Kalb-Ramond fields appears in all known string
theories. However, the spectrum of the $D=4$ \cite{nieu} and $D=3$ \cite
{witten} compactified theory from $D=10$ supergravity, contains the massive
antisymmetric tensor fields. Thus, since supersymmetry places severe
constraints on the ground state and the mass spectrum of the excitations,
supersymmetric mechanisms of mass generation are of considerable importance.

The purpose of this letter is twofold. First we construct an $N=1$ $-D=4$
superspace version of the $U(1)$ BF model. By means of a dimensional
reduction procedure, we obtain a massive antisymmetric tensor field into a $%
N=2-D=3$ supersymmetric topological massive gauge invariant theory. In
contrast to several works on $D=3$ BF models, we have considered here a
topological term which involves a KB and a pseudoscalar field with
derivative coupling. Secondly, we have addressed a $N=1$ superspace
mechanism to generate mass for Kalb-Ramond field without loss of gauge
invariance. Actually, this mechanism is a superspace version of the
topological massive formulation of Deser, Jackiw and Templeton \cite{jackiw}%
. On the other hand, an alternative model with an explicit mass breaking
term is constructed in $N=1$ superspace and a supersymmetric version of the
St\"{u}ckelberg transformation \cite{stuck} is used to restore the gauge
invariance of the model.

\section{The $N=1-D=4$ Extended BF Model}

Let us begin by introducing the $N=1-D=4$ supersymmetric BF extended model.
For extended we mean that we include mass terms for the Kalb-Ramond field.
This mass term will be introduced here for later comparison to the
tridimensional case. Actually, this construction can be seen as a superspace
and Abelian version of the so called BF-Yang-Mills models \cite{martellini}.

As our basic superfield action we take\footnote{%
Our spinorial notations and other conventions follow ref. \cite{superspace}.}

\begin{equation}
S_{BF}^{SS}=\frac 18\int d^4x\{-i\kappa [\int d^2\theta B^\alpha W_\alpha
-\int d^2\overline{\theta }\overline{B}_{\dot{\alpha}}\overline{W}^{\dot{%
\alpha}}]+\frac{g^2}2[\int d^2\theta B^\alpha B_\alpha +\int d^2\overline{%
\theta }\overline{B}_{\dot{\alpha}}\overline{B}^{\dot{\alpha}}]\}\mbox{ }.
\label{1.1}
\end{equation}
where $W_\alpha $ is a spinor superfield-strenght, $B_\alpha $ is a chiral
spinor superfield, $\bar{D}_{\dot{\alpha}}B_\alpha =0$, $\kappa $ and $g$
are massive parameters. Their corresponding $\theta $-expansions are:

\begin{eqnarray}
W_\alpha (x,\theta ,\bar{\theta}) &=&4i\lambda _\alpha (x)-[4\delta _\alpha
^\beta D(x)+2i(\sigma ^\mu \bar{\sigma}^\nu )_\alpha ^\beta F_{\mu \nu
}(x)]\theta _\beta  \nonumber \\
&&+4\theta ^2\sigma _{\alpha \dot{\alpha}}^\mu \partial _\mu \bar{\lambda}^{%
\dot{\alpha}}  \label{1.2a}
\end{eqnarray}

\begin{equation}
B_\alpha (x,\theta ,\overline{\theta })=e^{i\theta \sigma ^\mu \overline{%
\theta }\partial _\mu }[i\psi _\alpha (x)+\theta ^\beta T_{\alpha \beta
}(x)+\theta \theta \xi _\alpha (x)]\mbox{ ,}  \label{1.2b}
\end{equation}
where

\begin{equation}
T_{\alpha \beta }=T_{(\alpha \beta )}+T_{[\alpha \beta ]}=-4i(\sigma ^{\mu
\nu })_{\alpha \beta }B_{\mu \nu }+2\varepsilon _{\alpha \beta }(M+iN)%
\mbox{
}.  \label{1.2c}
\end{equation}

Our conventions for supersymmetric covariant derivatives are

\begin{eqnarray}
D_\alpha &\equiv &\frac \partial {\partial \theta ^\alpha }+i\sigma _{\alpha
\dot{\alpha}}^\mu \bar{\theta}^{\dot{\alpha}}\partial _\mu  \nonumber \\
\bar{D}_{\dot{\alpha}} &\equiv &-\frac \partial {\partial \bar{\theta}^{\dot{%
\alpha}}}-i\theta ^\alpha \sigma _{\alpha \dot{\alpha}}^\mu \partial _\mu %
\mbox{ .}  \label{1.2d}
\end{eqnarray}

We call attention for the electromagnetic field-strenght and the
antisymmetric gauge field which are contained in $W_\alpha $ and $B_\alpha $%
, respectively. In terms of the components fields, the action (\ref{1.1})
can be read as

\begin{eqnarray}
S &=&\int d^4x\{[-\frac{i\kappa }2\left( \xi \lambda -\bar{\xi}\bar{\lambda}%
\right) +\frac \kappa 2\left( \psi ^\alpha \sigma _{\alpha \dot{\alpha}}^\mu
\partial _\mu \bar{\lambda}^{\dot{\alpha}}+\bar{\psi}_{\dot{\alpha}}\left(
\bar{\sigma}^\mu \right) ^{\dot{\alpha}\alpha }\partial _\mu \lambda _\alpha
\right) +\frac \kappa 2B^{\mu \nu }\widetilde{F}_{\mu \nu }  \nonumber \\
&&-\kappa DN]+g^2[\frac 18\left( \psi \xi +\bar{\psi}\bar{\xi}\right) +\frac
12B^{\mu \nu }B_{\mu \nu }-\frac 12\left( M^2+N^2\right) ]\}  \nonumber \\
&=&\int d^4x[(\frac{i\kappa }2\bar{\Xi}\gamma ^5\Lambda +\frac \kappa 2\bar{%
\Psi}\gamma ^\mu \partial _\mu \Lambda +\frac \kappa 2B^{\mu \nu }\widetilde{%
F}_{\mu \nu }-\kappa DN)  \nonumber \\
&&+g^2(\frac 18\bar{\Psi}\Xi +\frac 12B^{\mu \nu }B_{\mu \nu }-\frac
12\left( M^2+N^2\right) )]\mbox{ }.  \label{comp}
\end{eqnarray}

In the last equality above, the fermionic fields have been organized as
four-component Majorana spinors as follows

\begin{equation}
\Xi =\left(
\begin{array}{c}
\xi _\alpha \\
\bar{\xi}^{\dot{\alpha}}
\end{array}
\right) \mbox{ };\mbox{ }\Lambda =\left(
\begin{array}{c}
\lambda _\alpha \\
\bar{\lambda}^{\dot{\alpha}}
\end{array}
\right) \mbox{ };\mbox{ }\Psi =\left(
\begin{array}{c}
\psi _\alpha \\
\bar{\psi}^{\dot{\alpha}}
\end{array}
\right) \mbox{ },  \label{1.4}
\end{equation}
and we denote the dual field-strenght defining $\widetilde{F}_{\mu \nu
}\equiv \frac 12\varepsilon _{\mu \nu \alpha \beta }F^{\alpha \beta }$.
Furthermore, we use the following identities

\begin{eqnarray}
\bar{\Psi}\Lambda &=&\bar{\psi}\bar{\lambda}+\psi \lambda  \nonumber \\
\bar{\Psi}\gamma ^5\Lambda &=&\bar{\psi}\bar{\lambda}-\psi \lambda  \nonumber
\\
\bar{\Psi}\gamma ^\mu \Lambda &=&\psi \sigma ^\mu \bar{\lambda}+\bar{\psi}%
\bar{\sigma}^\mu \lambda \mbox{ .}  \label{1.4a}
\end{eqnarray}

The superfield action (\ref{1.1}) is a particular case of the action
proposed in ref. \cite{marcony}. However, a point of difference must be
noted. In contrast with \cite{marcony}, we have not considered coupling with
matter fields and a propagation term for the gauge fields. On the other
hand, our supespace BF term was constructed in a distinct and simpler way. A
quite similar construction was introduced by Clark {\it et al.} \cite{clark}.

The off-diagonal mass term $\xi \lambda $ (or $\bar{\Xi}\gamma ^5\Lambda $)
has been shown by Brooks and Gates, Jr. \cite{brooks} in the context of
super-Yang-Mills theory. Note that the identity

\begin{equation}
\gamma _5\sigma ^{\mu \nu }=\frac i2\varepsilon _{\mu \nu \alpha \beta
}\sigma ^{\alpha \beta }  \label{civita}
\end{equation}
reveals a connection between the topological behaviour denoted by the
Levi-Civita tensor $\varepsilon _{\mu \nu \alpha \beta },$ and the
pseudo-escalar $\gamma _5.$

So, it is worthwhile to mention that this term has topological origin and it
can be seen as a fermionic counterpart of the BF term. In our opinion, this
fermionic mass term deserves more attention and will be investigated
elsewhere.

\section{The $N=2-D=3$ Topological Model}

As it is well known, the BF model in $D=3$ consists in a one form field ($%
"B" $ field) and one form gauge field $A$. So, the Chern-Simons term is
simply the identification of $B$ and $A$. However, as has been shown in ref.
\cite{dedit}, after dimensional reduction of the four dimensional BF model,
an interesting additional term arises, namely, a topological term which
involves a 2-form and a 0-form. We will call it a $B\varphi $ term. A quite
similar model was presented in a Yang-Mills version by Del Cima {\it et. al}%
. \cite{osvaldo}, and its finiteness was proved in the framework of
algebraic renormalization.

Following the procedure of ref. \cite{dedit}, we will carry out a
dimensional reduction in the bosonic sector of (\ref{comp}). Dimensional
reduction is usually done by expanding the fields in normal modes
corresponding to the compactified extra dimensions, and integrating out the
extra dimensions. This approach is very useful in dual models and
superstrings \cite{scherk}. Here, however, we only consider the fields in
higher dimensions to be independent of the extra dimensions. In this case,
we assume that our fields are independent of the extra coordinate $x_3.$

Therefore, after dimensional reduction, the bosonic sector of (\ref{comp})
can be written as

\begin{eqnarray}
S_{bos.} &=&\int d^3x\{[\kappa \varepsilon _{\mu \alpha \beta }V^\mu
F^{\alpha \beta }+\kappa \varepsilon _{\mu \nu \alpha }B^{\mu \nu }\partial
^\alpha \varphi -\kappa DN]  \nonumber \\
&&+g^2[\frac 12B^{\mu \nu }B_{\mu \nu }-V^\mu V_\mu -\frac 12\left(
M^2+N^2\right) ]\}\mbox{ }.  \label{1.5}
\end{eqnarray}
Notice that the first term in r.h.s. of (\ref{1.5}) can be transformed in
the Chern-Simons term if we identify $V^\mu \equiv A^\mu $. The second one
is the so called $B\varphi $ term.

Now let us proceed to the dimensional reduction of the fermionic sector of
the model. First, note that the Lorentz group in three dimensions is $%
SL(2,R) $ rather than $SL(2,C)$ in $D=4$. Therefore, Weyl spinors with four
degrees of freedom will be mapped into Dirac spinors\footnote{%
For details about spinorial dimensional reduction, we suggest refs. \cite
{refa5} and \cite{refa6}.}. So the correct associations keeping the degrees
of freedom are sketched as

\begin{eqnarray}
\Xi &=&\left(
\begin{array}{c}
\xi _\alpha \\
\bar{\xi}^{\dot{\alpha}}
\end{array}
\right) \rightarrow \Xi _{\pm }=\xi _\alpha \pm i\tau _\alpha  \nonumber \\
\Lambda &=&\left(
\begin{array}{c}
\lambda _\alpha \\
\bar{\lambda}^{\dot{\alpha}}
\end{array}
\right) \rightarrow \Lambda _{\pm }=\lambda _\alpha \pm i\rho _\alpha
\nonumber \\
\Psi &=&\left(
\begin{array}{c}
\psi _\alpha \\
\bar{\psi}^{\dot{\alpha}}
\end{array}
\right) \rightarrow \Psi _{\pm }=\psi _\alpha \pm i\chi _\alpha \mbox{ }.
\label{1.6}
\end{eqnarray}

From ($\ref{1.6}$), we find that

\begin{eqnarray}
\Psi \bar{\Xi} &\rightarrow &\frac 12\left( \Psi _{+}\Xi _{-}+\Psi _{-}\Xi
_{+}\right)  \nonumber \\
\bar{\Psi}\gamma ^\mu \partial _\mu \Lambda &\rightarrow &\frac 12(\Psi
_{+}\gamma ^{\widehat{\mu }}\partial _{\widehat{\mu }}\Lambda _{-}+\Psi
_{-}\gamma ^{\widehat{\mu }}\partial _{\widehat{\mu }}\Lambda _{+})
\nonumber \\
\Xi \gamma ^5\Lambda &\rightarrow &\frac 12(\Xi _{+}\Lambda _{+}+\Xi
_{-}\Lambda _{-})\mbox{ }.  \label{1.7}
\end{eqnarray}
where $hatted$ index means three-dimensional space-time.

Thus, the dimensionally reduced fermionic sector of (\ref{comp}) may be
written

\begin{eqnarray}
S_{ferm.} &=&\int d^3x\{\frac{i\kappa }4\left( \Xi _{+}\Lambda _{+}+\Xi
_{-}\Lambda _{-}\right) +\frac \kappa 4(\Psi _{+}\gamma ^{\widehat{\mu }%
}\partial _{\widehat{\mu }}\Lambda _{-}+\Psi _{-}\gamma ^{\widehat{\mu }%
}\partial _{\widehat{\mu }}\Lambda _{+})  \nonumber \\
&&+\frac{g^2}{16}\left( \Psi _{+}\Xi _{-}+\Psi _{-}\Xi _{+}\right) \}%
\mbox{
.}  \label{1.8}
\end{eqnarray}

The action $S=S_{bos.}+S_{ferm.}$ is invariant under the following
supersymmetry transformations

\begin{eqnarray}
\delta \lambda _\alpha &=&-iD\eta _\alpha -\left( \sigma ^\mu \sigma ^\nu
\right) _\alpha ^\beta \eta _\beta F_{\mu \nu }  \nonumber \\
\delta \rho _\alpha &=&iD\zeta _\alpha -\left( \sigma ^\mu \sigma ^\nu
\right) _\alpha ^\beta \zeta _\beta F_{\mu \nu }  \nonumber \\
\delta F^{\mu \nu } &=&i\partial ^\mu \left( \eta \sigma ^\nu \rho -\lambda
\sigma ^\nu \zeta \right) -i\partial ^\nu \left( \eta \sigma ^\mu \rho
-\lambda \sigma ^\mu \zeta \right)  \nonumber \\
\delta D &=&\partial _\mu \left( -\eta \sigma ^\mu \rho +\lambda \sigma ^\mu
\zeta \right)  \label{1.9a}
\end{eqnarray}

\begin{eqnarray}
\delta \left( \psi _\alpha \pm i\chi _\alpha \right) &=&\delta \Psi _{\pm
}=i\eta ^\beta \widetilde{T}_{\beta \alpha }\pm \zeta ^\beta \widetilde{T}%
_{\beta \alpha }  \nonumber \\
\delta \widetilde{T}_{\beta \alpha } &=&-\eta _\beta \xi _\alpha +\zeta
^\lambda \sigma _{\beta \lambda }^\mu \partial _\mu \psi _\alpha  \nonumber
\\
\delta \left( \xi _\alpha \pm i\tau _\alpha \right) &=&\delta \Xi _{\pm
}=-i\zeta _\lambda \left( \sigma ^\mu \right) ^{\lambda \beta }T_{\beta
\alpha }\mp \eta _\lambda \left( \bar{\sigma}^\mu \right) ^{\beta \lambda
}T_{\beta \alpha }\mbox{ ,}  \label{1.9b}
\end{eqnarray}
where $\eta $ and $\zeta $ are supersymmetric parameters, which indicates
that we have two supersymmetries in the aforementioned action.

\section{Remarks on Some 3D Supersymmetric Models and St\"{u}ckelberg
Formulation}

From the two topological terms introduced in (\ref{1.5}) we can set up two
supersymmetric models. The first one, which involves a two and a zero form,
can be expressed as

\begin{equation}
S=\int d^3xd^2\theta (D^\alpha \Phi B_\alpha +\frac 12g^2B^\alpha B_\alpha )%
\mbox{ },  \label{1.10}
\end{equation}
where $B_\alpha $ and $\Phi $ are spinor and real scalar superfields, which
are defined by projection as

\begin{eqnarray}
B_\alpha &\mid &=\chi _\alpha  \nonumber \\
D_{(\beta }B_{\alpha )} &\mid &=2iM_{\beta \alpha }=M_{\alpha \beta }=B^{\mu
\nu }\left( \sigma _{\mu \nu }\right) _{\alpha \beta }  \nonumber \\
D^\alpha B_\alpha &\mid &=2N  \nonumber \\
D^\beta D_\alpha B_\beta &\mid &=2\omega _\alpha  \label{1.11a}
\end{eqnarray}
and

\begin{eqnarray}
\Phi &\mid &=\varphi  \nonumber \\
D_\alpha \Phi &\mid &=\psi _\alpha  \nonumber \\
D^2\Phi &\mid &=F\mbox{ }.  \label{1.11b}
\end{eqnarray}

Here the supersymmetry covariant derivative is given by $D_\alpha =\partial
_\alpha +i\theta ^\beta \partial _{\alpha \beta }$ . So, in terms of
components fields, the action (\ref{1.10}) becomes

\begin{eqnarray}
S &=&\int d^3x[\left( \kappa \partial ^{\alpha \beta }\varphi M_{\beta
\alpha }+2\kappa \psi ^\alpha \omega _\alpha -2\kappa FN\right)  \nonumber \\
&&+\frac 12g^2\left( 4\omega ^\alpha \chi _\alpha +2i\chi _\alpha \partial
^{\beta \alpha }\chi _\beta +M^{\beta \alpha }M_{\alpha \beta }+2N^2\right) ]%
\mbox{ }.  \label{1.12}
\end{eqnarray}

Starting from the definitions of two spinor superfields given by

\begin{eqnarray}
\Lambda _\alpha &\mid &=\xi _\alpha  \nonumber \\
D_{(\beta }\Lambda _{\alpha )} &\mid &=2iV_{\beta \alpha }  \nonumber \\
D^\alpha \Lambda _\alpha &\mid &=2G  \nonumber \\
D^\beta D_\alpha \Lambda _\beta &\mid &=2\rho _\alpha  \label{1.13a}
\end{eqnarray}
and

\begin{eqnarray}
W_\alpha &\mid &=\lambda _\alpha  \nonumber \\
D_\alpha W_\beta &\mid &=f_{\alpha \beta }\mbox{ },  \label{1.13b}
\end{eqnarray}
where

\begin{equation}
V_{\beta \alpha }\equiv V^\mu \left( \widetilde{\sigma }_\mu \right) _{\beta
\alpha }\mbox{ };\mbox{ }f_{\alpha \beta }\equiv \left( \tilde{\sigma}_\mu
\right) _{\alpha \beta }f^\mu \mbox{ };\mbox{ }f^\mu =-\frac i2\varepsilon
^{\mu \nu \rho }F_{\nu \rho }\mbox{ },  \label{1.13c}
\end{equation}
we can propose another supersymmetric action, now involving two 1-forms,
namely

\begin{eqnarray}
S &=&\int d^3xd^2\theta (\Lambda ^\alpha W_\alpha -g^2\Lambda ^\alpha
\Lambda _\alpha )  \nonumber \\
&=&\int d^3x[\left( 2\rho ^\alpha \lambda _\alpha -iV^{\alpha \beta
}f_{\beta \alpha }\right)  \nonumber \\
&&-g^2\left( 4\rho ^\alpha \omega _\alpha +2i\xi _\alpha \partial ^{\beta
\alpha }\xi _\beta +V^{\beta \alpha }V_{\beta \alpha }+2G^2\right) ]\mbox{ .}
\label{1.14}
\end{eqnarray}

It is easy to see that the superspace actions (\ref{1.10}) and (\ref{1.14})
are not invariant under the following gauge transformations

\begin{eqnarray}
\delta B^\alpha &=&D^\beta D^\alpha \Pi _\beta  \nonumber \\
\delta \Phi &=&0  \label{1.15a}
\end{eqnarray}

\begin{eqnarray}
\delta \Lambda ^\alpha &=&D^\alpha \Omega  \nonumber \\
\delta W^\alpha &=&0\mbox{ }.  \label{1.15b}
\end{eqnarray}

However, if we reparameterize $\Lambda ^\alpha $ and $B^\alpha $
through introduction of the St\"{u}ckelberg superfields\footnote{%
For historical reasons, it is important to cite here the first
work, to the best of our knowledge, in the framework of
supersymmetric St\"{u}ckelberg formalism, namely ref.
\cite{delbourgo}.} $\Theta $ and $\Sigma _\alpha $ such that

\begin{eqnarray}
\Lambda ^\alpha &\rightarrow &\left( \Lambda ^\alpha \right) ^{\prime
}=\Lambda ^\alpha +\frac 1gD^\alpha \Theta  \nonumber \\
B^\alpha &\rightarrow &\left( B^\alpha \right) ^{\prime }=B^\alpha +D^\beta
D^\alpha \Pi _\beta \mbox{ },  \label{1.16}
\end{eqnarray}
and imposing that $\Theta $ and $\Sigma _\alpha $ transform like

\begin{eqnarray}
\delta \Theta &=&-g\Omega  \nonumber \\
\delta \Sigma ^\beta &=&-\Pi ^\beta \mbox{ ,}  \label{1.17}
\end{eqnarray}
we ensure gauge invariance for that superactions.

We remark that integrating out the superfield $B_\alpha $ in the equation (%
\ref{1.10}) we arrive at a supersymmetric Klein-Gordon action and, if we do
the same for $\Lambda _\alpha $ in (\ref{1.14}), we obtain a Maxwell
superaction. Observe that both these relations may be understood as two
duality tranformations. We recall here that an analogous connection in $4D$
pure bosonic BF-theory was viewed as a perturbative expansion in the
coupling $g$ around the topological pure BF theory \cite{fucito}. Thereupon,
it may be interesting to perform a similar investigation in the framework of
action (\ref{1.10}).

\section{$N=1$ Superspace Topological Mass Generation}

In order to show the topological mass generation for the Kalb-Ramond two
form field, we will construct a variation from the model (\ref{1.10}), by
introducing the propagation term for it. Before that, for ilustration
purpose, we quote the bosonic action introduced in ref. \cite{dedit}:

\begin{equation}
S=\int d^3x\left[ \frac 16H_{\mu \nu \rho }H^{\mu \nu \rho }+k\epsilon _{\mu
\nu \rho }B^{\mu \nu }\partial ^\rho \phi +\frac 12\partial _\mu \phi
\partial ^\mu \phi \right] \mbox{ },  \label{1.18}
\end{equation}
where $H_{\mu \nu \rho }$, a three form field-strength of the $B^{\mu \nu }$
field, is defined as

\begin{equation}
H_{\mu \nu \rho }=\partial _{[\mu }B_{\mu \rho ]}=\partial _\mu B_{\nu \rho
}+\partial _\nu B_{\rho \mu }+\partial _\rho B_{\mu \nu }\mbox{ }.
\label{1.19}
\end{equation}

The $N=1$ superspace construction of the supersymmetric version of (\ref
{1.18}) proceeds as follows. First, we introduce a scalar superfield $G$
defined by

\begin{equation}
G=-D^\alpha B_\alpha \mbox{ },  \label{1.20}
\end{equation}
where $B_\alpha $ is the super-Kalb-Ramond field defined in (\ref{1.11a}).
Then, after looking the expression (\ref{1.10}), we find the action

\begin{equation}
S=\int d^3xd^2\theta [-\frac 12\left( D^\alpha G^2\right) +kB^\alpha
D_\alpha \Phi -\frac 12D^\alpha \Phi D_\alpha \Phi ]\mbox{ }.  \label{1.21}
\end{equation}

Now it is straightforward to show that the topological term $kB^\alpha
D_\alpha \Phi $ gives rise to a mass term for the super-Kalb-Ramond field.
The equation of motion associated with $\Phi $ is,
\begin{equation}
D^\alpha \left( kB_\alpha -D_\alpha \Phi \right) =0\mbox{ }.  \label{1.22}
\end{equation}
Consequently,

\begin{equation}
kB_\alpha -D_\alpha \Phi ={\cal C}\mbox{ .}  \label{1.23}
\end{equation}
Since that the constant ${\cal C}$ can be absorbed by $B_\alpha $, we
conclude that

\begin{equation}
kB_\alpha -D_\alpha \Phi =0\mbox{ .}  \label{1.24}
\end{equation}

Therefore the original action (\ref{1.21}) can be rewritten as

\begin{equation}
S=\int d^3xd^2\theta [\left( D^\alpha G^2\right) +\frac 12k^2B^\alpha
B_\alpha ]\mbox{ }.  \label{1.25}
\end{equation}

This exhibits a topological mechanism of mass generation for the Kalb-Ramond
field. Naturally, the topological mass terms arise due to the coupling of
the $B_\alpha $ and $\Phi $ superfields. In other words, this mass term
results of the breakdown of the gauge invariance (\ref{1.15a}).

Incidentally let us mention a possible equivalence similar to that between
massive topologically and self-dual theories in $D=3$ \cite{jackiw}. Indeed,
starting from (\ref{1.10}), we can construct an action by introduction of a
mass term for the superfield $\Phi ,$ namely

\begin{equation}
S=\int d^3xd^2\theta (D^\alpha \Phi B_\alpha +\frac 12g^2B^\alpha B_\alpha
+m\Phi ^2)\mbox{ }.  \label{1.26}
\end{equation}

It is easy to see that the equations of motion of (\ref{1.26}) and (\ref
{1.21}) are equivalent. So, the action (\ref{1.26}) can be considered
locally equivalent to action (\ref{1.21}). On the other hand, it would be
interesting to investigate if this equivalence is preserved at quantum level.

\section{Conclusions}

In this letter, we have constructed an $N=1-$ $D=3$ superspace action for a
model involving an antisymmetric gauge field. Our main point is a
topological term that consists in a coupling of this 2-form field and a
scalar field. To the best of our knowledge, in the form presented here, this
model is completely new in the literature. A similar approach, but involving
a 3-form and a scalar fields in $N=1-D=4$, was introduced in ref. \cite
{khoudeir}.

Starting from the so called $B\wedge F$ model in $N=1-D=4$ superspace, we
carried out a dimensional reduction to the three-dimensional space-time, in
order to obtain our basic model. The superspace construction for the $%
B\wedge F$ is known, but we point out the appearance of a fermionic
counterpart of the $B\wedge F$ term.

We have introduced two massive gauge invariant models for an antisymmetric
tensor field into a $N=1-D=3$ superspace. In the first, we resort to the
St\"{u}ckelberg formalism and in the other, we construct an abelian
topologically massive theory, and a topologically generated mass for the
Kalb-Ramon superfield is exhibited. An equivalence of both massive models is
suggested. Furthermore, a component field analysis is performed, showing a
second supersymmetry in the model.

\vspace{0.3in} \centerline{\bf ACKNOWLEDGMENTS}

We wish to thank M. S. Cunha for helpful discussions. Prof. J. E. Moreira is
acknowledged by a critical reading of the manuscript. This work was
supported in part by Conselho Nacional de Desenvolvimento Cient\'{\i }fico e
Tecnol\'{o}gico-CNPq and Funda\c{c}\~{a}o Cearense de Amparo \`{a}
Pesquisa-FUNCAP.


\end{document}